\documentclass[12pt]{iopart}
\usepackage{listings}
\usepackage{hyperref}
\usepackage{graphicx}
\usepackage[utf8]{inputenc}
\usepackage{tikz}
\usepackage{pgfplots}
\usepackage[labelfont=bf]{caption}
\usepackage{subfig}


\begin{document}

\title{ROOT I/O compression algorithms and their performance impact within Run 3}

\author{Oksana Shadura [1], Brian Paul Bockelman  [2]}
\address{[1] University of Nebraska-Lincoln, USA, [2] Morgridge Institute for Research, USA}
\ead{[1] oksana.shadura@cern.ch, [2] bbockelman@morgridge.org}
\vspace{10pt}

\begin{abstract}

The LHC’s Run3 will push the envelope on data-intensive workflows and, since at the lowest level this data is managed using the ROOT software framework, preparations for managing this data are starting already. At the beginning of LHC Run 1, all ROOT data was compressed with the ZLIB algorithm; since then, ROOT has added support for additional algorithms such as LZMA and LZ4, each with unique strengths.   This work must continue as industry introduces new techniques - ROOT can benefit saving disk space or reducing the I/O and bandwidth for online and offline needs of experiments by introducing better compression algorithms. In addition to alternate algorithms, we have been exploring alternate techniques to improve parallelism and apply pre-conditioners to the serialized data.

We have performed a survey of the performance of the new compression techniques. Our survey includes various use cases of data compression of ROOT files provided by different LHC experiments. We also provide insight into solutions applied to resolve bottlenecks in compression algorithms, resulting in  improved ROOT performance.
\end{abstract}

\section{Introduction}

The need for data compression within the HEP community has grown significantly as the amount of data collected, transmitted, and stored has increased throughout the LHC era. Over the next two years, the community is preparing for Run 3 (2020 - 2022).  During these years, the LHC will increase both energy levels and instantaneous luminosity, increasing the size of each event and putting further pressure on the storage systems.

As the community needs to push boundaries, we observe more interest in specializing the lossless compression algorithms used by ROOT \cite{root} (in addition to more physics-specific lossy compression).  Historically, ZLIB has been used for all HEP data as it is a general-purpose algorithm.  However, the problem faced by production (high compression ratio needed, significant CPU per event available) is very different from most analysis (less constrainted by total volume but little per-event CPU available); we have found these cases may match alternative, specialized algorithms better than the general-purpose ZLIB.

This paper is organized as followed: Section 2 gives overview of ROOT compression algorithms. It introduces, explains and compares performance of four different compression libraries: \textit{ZLIB}, \textit{ZLIB-CF}, \textit{LZ4} and \textit{ZSTD}. Section 3 describes results of each algorithm's benchmarking in ROOT.

\section{ROOT compression algorithms}

\begin{figure}[h]
\centering
\includegraphics[width=0.8\linewidth]{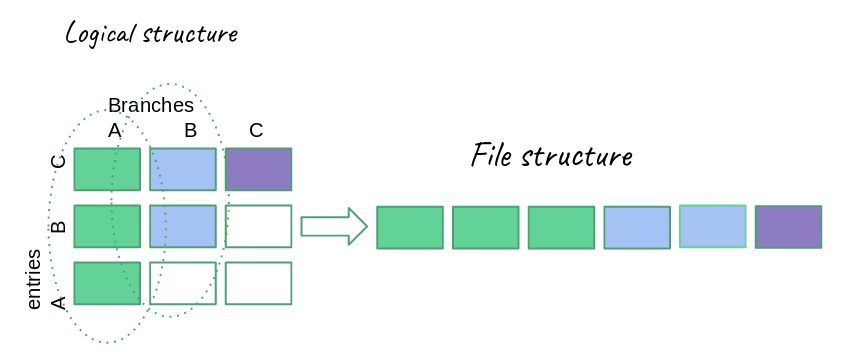}
\caption{ROOT I/O schema.  Data laid out logically into ``branches" and ``entries" (corresponding to columns and rows in a table) are serialized, column-wise, into buffers.  These buffers are then compressed and written into disk as part of a ROOT file.  The structure containing the buffers are referred to as `baskets'.}
\label{fig:rootio}
\end{figure}

Reading input in ROOT I/O subsystem consists primarily of decompression and deserialization operations. Compression and decompression are significant for ROOT I/O (see the Figure \ref{fig:rootio}) and are a ROOT core functionality. The compressed baskets entries (green, blue and purple entries on the Figure \ref{fig:rootio} and a part of the logical structure of a ROOT file) present a number of advanced compression or decompression possibilities such as simultaneous read and decompression for the multiple physics events.

The available set of compression algorithms in ROOT 6.18.00 are:
\begin{enumerate}
    \item \textit{ZLIB} - a LZ77-based compressor with Huffman coding \cite{zlib}.
    \item \textit{LZMA} - one of the LZ77 family of the compressors, which has significantly larger dictionary sizes compared to ZLIB compression algorithm and has more complex encoding techniques, such as use of a range encoder (using a complex model for probability-based prediction) \cite{lzma}.
    \item \textit{Custom ROOT compression algorithm}.  This ZLIB-like compression algorithm, dating back to the 1990's, is used only for ROOT backward compatibility.
    \item \textit{LZ4} - a LZ77-based algorithm with a fixed, byte-oriented encoding and without Huffman coding pass \cite{lz4} \cite{brianzhe}.
\end{enumerate}

Further, we have developed a test integration - not part of any ROOT release - of \textit{ZSTD}, a LZ77-based algorithm supporting dictionaries, a large search window, and a entropy coding stage, using fast Finite State Entropy (tANS) or Huffman coding \cite{zstd}.

Each algorithm is exposed with a single tunable parameter (which ROOT refers to as ``compression level") that allows the user to choose between faster compression speeds (``level 1") and improved compression ratio with lower compression speed - (``level 9"). Additionally, compression level 0 indicates all compression has been disabled.  To test compression and decompression performance and compression ratio, we utilize a simple test case of an artificially-generated ROOT tree with 2,000 events.

\begin{figure}[h]
\centering
\includegraphics[width=0.8\linewidth]{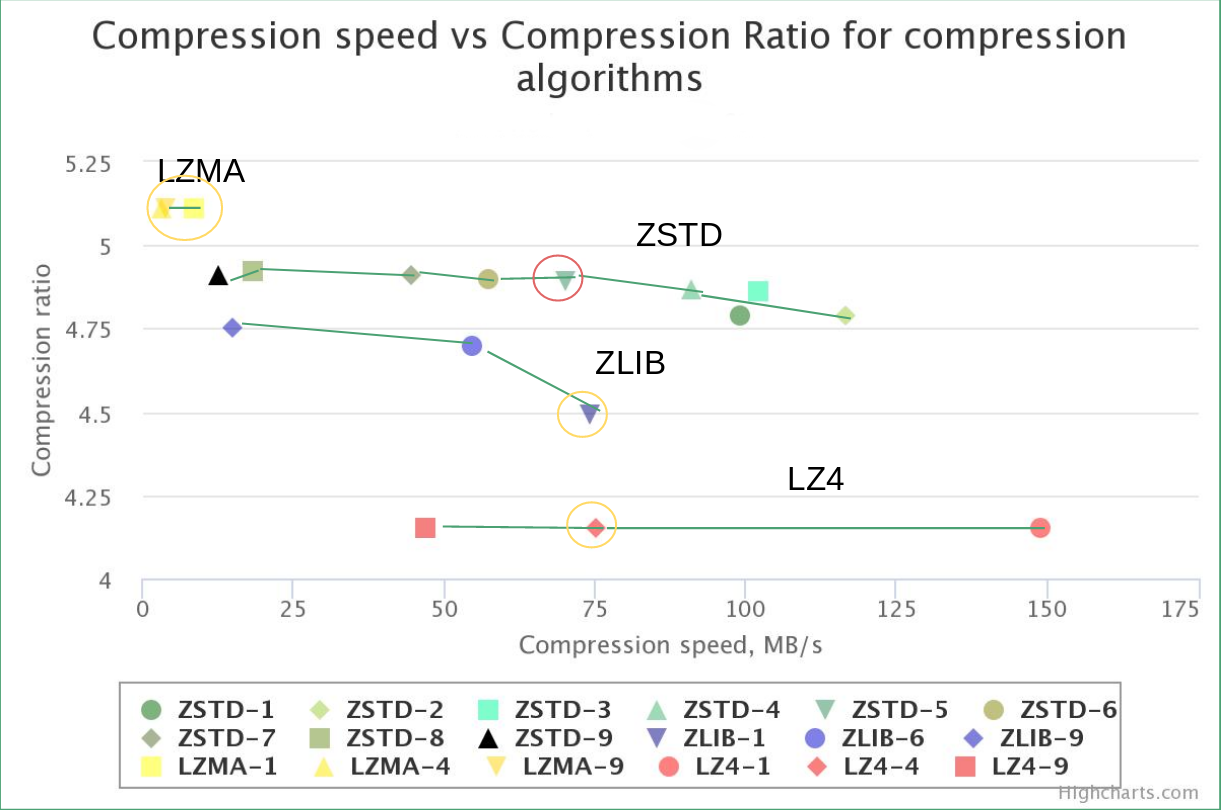}
\caption{Diagram demonstrating the performance of various compression algorithms in ROOT.  The x-axis demonstrates the overall compression ratio for the test file while the y-axis shows the compression speed.  Each data point is a combination of an algorithm and a compression level.  This test was run utilizing a host with a Haswell-class Intel processor and a SSD.}
\label{fig:compression}
\end{figure}

\begin{figure}[h]
\centering
\includegraphics[width=0.9\linewidth]{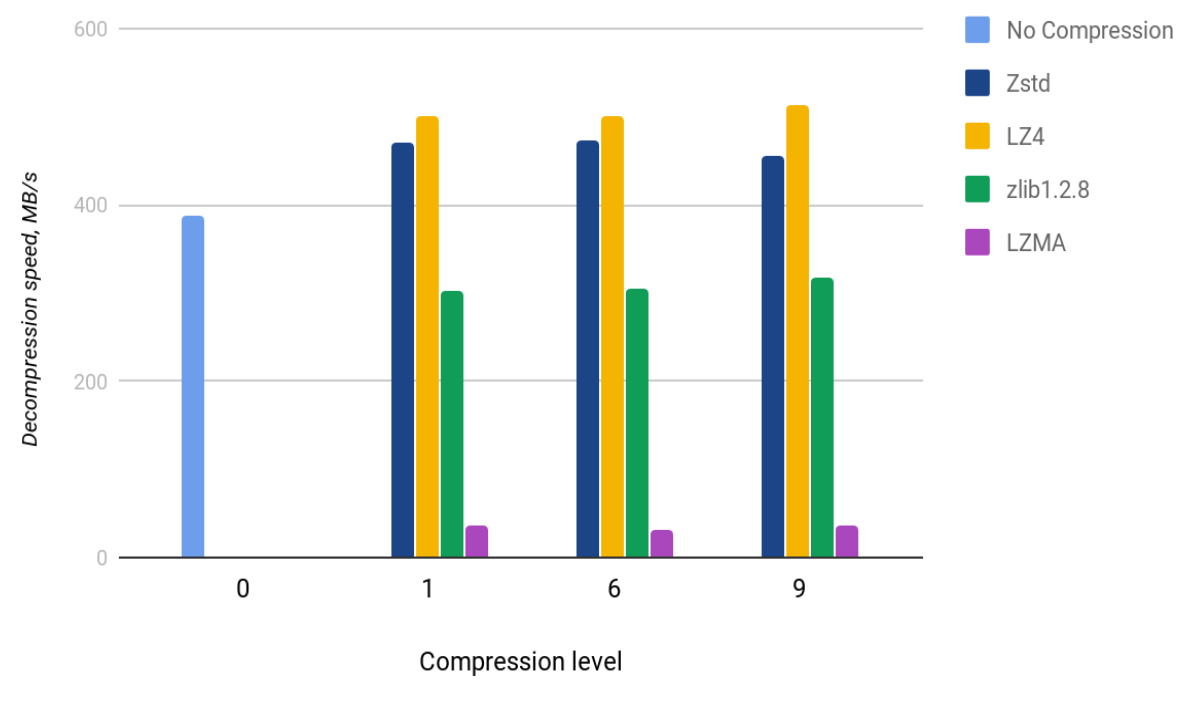}
\caption{Diagram demonstrating the decompression speed of reading a file with ROOT, varied by by algorithm and compression level of the input file.  Note the decompression speed is primarily a function of the compression algorithm and not level (compression levels 0, 1, 6, and 9 are illustrated).}
\label{fig:decompression}
\end{figure}

\subsection{Cloudflare Zlib}

The \textit{ZLIB} compression algorithm has had a venerable reference implementation maintained by Mark Adler for a wide variety of platforms.  More recently, CloudFlare \cite{zlib-cf} has maintained an alternate implementation (referred to as \textit{CF-ZLIB}), focused on performance rather than wide compatibility, for Intel/ARMv8 processors.

The classic ZLIB (also referred to as \textit{deflate}) compressed data format consists of a series of blocks, corresponding to successive blocks of input data. Each block is compressed using a combination of the LZ77 algorithm and Huffman coding. Performance hotspots for \textit{ZLIB} include checksum generations for both the \textit{adler32} and \textit{crc32} hash generation (while CloudFlare utilizes crc32, ROOT utilizes adler32).
 
For systems supporting Intel SSE4.2 and AVX (as well as NEON vector extensions in case of ARMv8), the existence \textit{crc32}-specific processor instructions improve overall checksum speed and other intrinsics speedup hash calculation. Both factors results in an overall speed improvements of hashing operations in the compression algorithm.  We used a similar approach to improve the speed of the \textit{adler32} checksum generation widely used in ZLIB.  For adler32, the important intrinsic function is \textit{\_mm\_sad\_epu8()} (which utilizes SSE4.2-provided instructions), which is used for byte summations and for accumulation of the byte sums uses SSE-style shuffle-adds.
 
 \textit{ZLIB} uses a hash table for look up of the location of previous occurrences of a given string in the sliding history window. For the fast compression levels (1-5), \textit{CF-ZLIB} hashes all bytes as quadruplets, resulting in a hash map entry for matches of four elements. This reduces the size of the hash map, while the reference \textit{ZLIB} implementation calculates hash entries based on triplet.  On newer processors, quadruplet-based calculation can be done utilizing vector instructions resulting in significant speedup.

The reference \textit{ZLIB} code base, dating back to 1995, has been optimized over the years for a series for a set of old platforms and compilers. Some of these hand-written optimziations, such as excessive loop unrolling, were effective patterns in past, but no longer needed for modern processors; in fact, these optimizations cause an overall slowdown on newer processors and compilers. In \textit{ZLIB}, In \textit{CF-ZLIB}'s \textit{adler32} implementation, the loop unrolling was reduced from 16 to 8; similarly, the unrolling in the \textit{crc32} computation was reduced from 8 to 4.
 
\begin{figure}[!ht]
\centering
\includegraphics[width=0.7\linewidth]{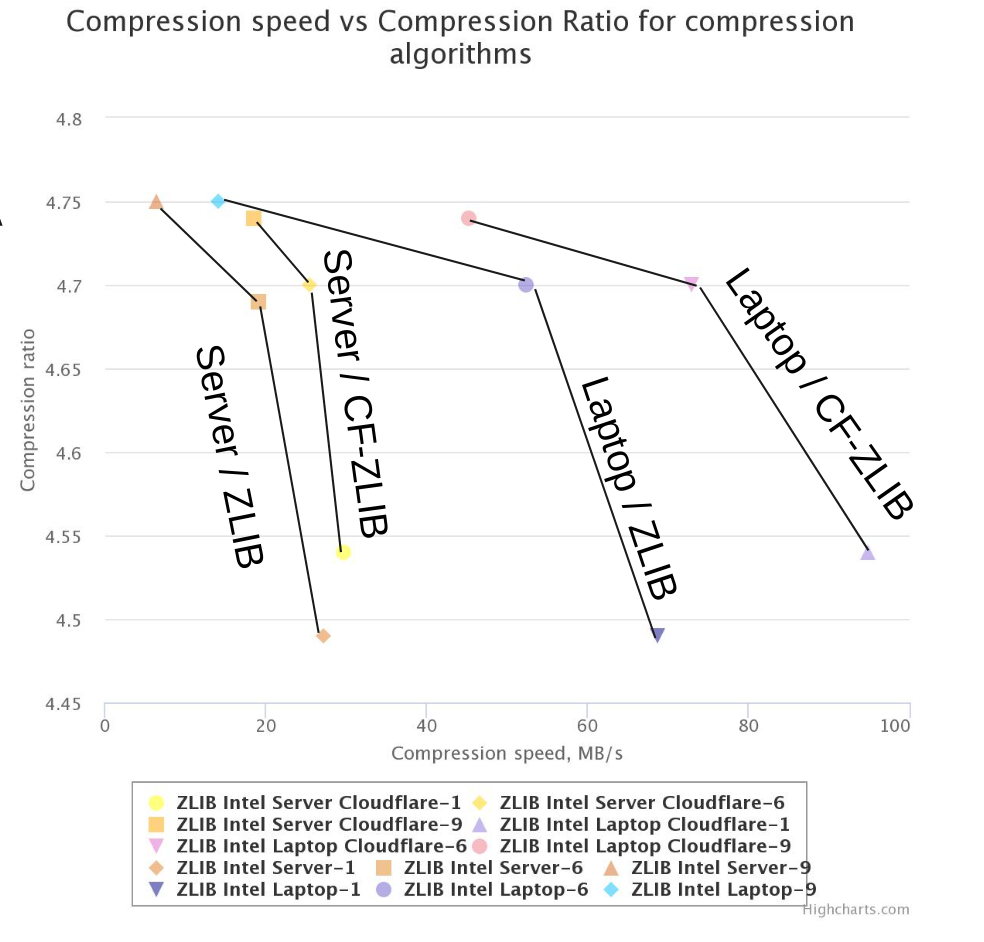}
\caption{Compression speed improvements from the CF-ZLIB patch set on a laptop-class and server-class platform.  Note that, on the laptop platform, the speed of a high level of compression (CF-ZLIB-6) is now similar to the previous speed for the lowest compression level (ZLIB-1). }
\label{fig:cflaptop}
\end{figure}

\begin{figure}[!ht]
\centering
\includegraphics[width=0.7\linewidth]{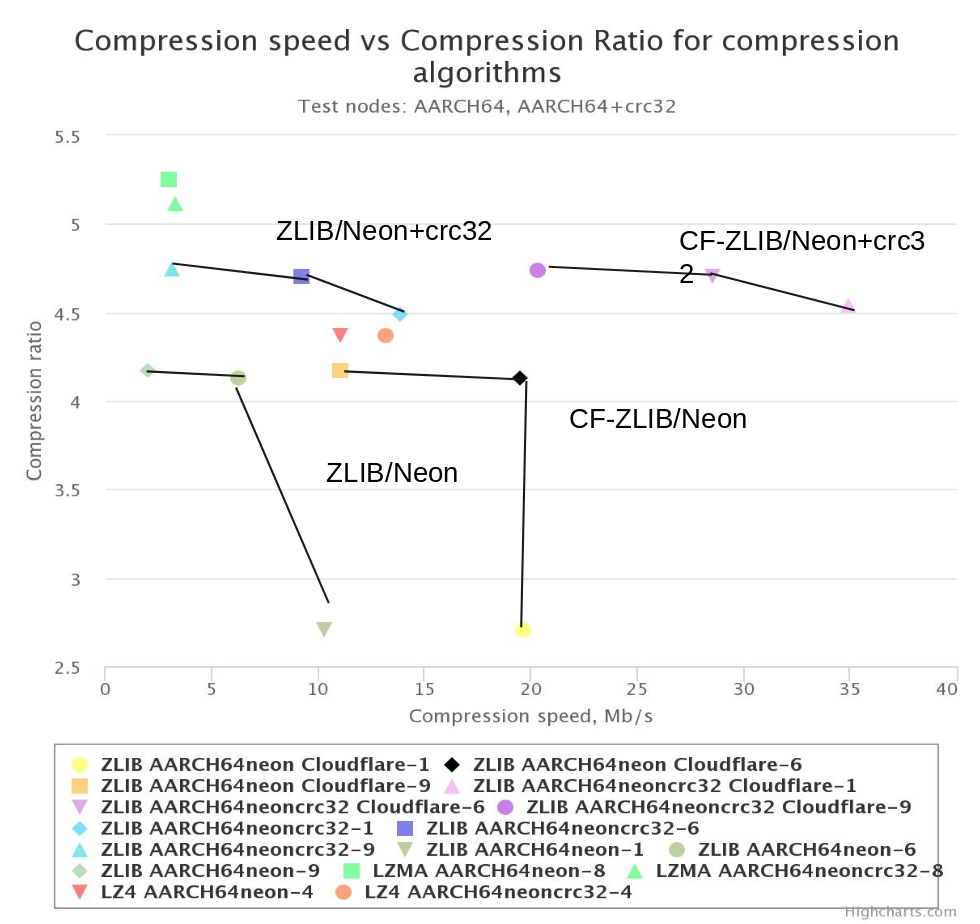}
\caption{Compression speed improvements from the CF-ZLIB patch set for two different configurations: with hardware implementation of crc32 calculations and without.}
\label{fig:cfneon}
\end{figure}

For ROOT 6.18.00, we have contributed a set of patches to switch the ZLIB algorithm from  the reference \textit{ZLIB} implementation to the \textit{CF-ZLIB} one.  The performance changes are summarized in Figures \ref{fig:cflaptop} and \ref{fig:cfneon}. The performance results in Figure \ref{fig:cflaptop} were measured on an Intel Core  i7 Haswell laptop-class processor equipped with SSD and Intel Haswell  Intel Xeon E5 v3 server-class processor; both support the Intel SSE 4.2 and AVX2 instruction sets.
 
For the ARM aarch64 platform (labeled ``AARCH64+CRC32" in Figure \ref{fig:cfneon}), we used HiSilicon's Hi1612 processor (Taishan 2180) which provides the Neon instruction set and a hardware implementation of the \textit{crc32} instructions.

The compression tests were based on \textit{Cloudflare ZLIB} \cite{zlib-cf-sources} with extensions from CMS \cite{zlib-cf-cms} (providing fallback compatibility on processors without SSE 4.2).  The code contributed to ROOT contained further improvements in terms of processor compatibility and performance.  Note that, due to the changes in the hashing function used for the Huffman coding pass, the compression ratios for \textit{CF-ZLIB} and \textit{ZLIB} vary slightly even at equivalent compression levels.

\subsection{LZ4}

\textit{LZ4} is a lossless compression algorithm focusing on providing the maximum compression and decompression speed. One of the key features is its extremely fast decompressor at all compression levels; the reference library contains both a high-speed compressor (\textit{LZ4}) and a a slower compressor which achieves higher compression ratios (\textit{LZ4-HC}).  The \textit{LZ4-HF} variant typically results in a 20\% improvement of compression ratio; see the comparison on Figure \ref{fig:compression}).  The \textit{LZ4} compression library with both compressors is available as of ROOT 6.14. 


One weakness of both \textit{LZ4} compression variants revealed during testing was the relatively poor compression ratio on certain ROOT files.  In ROOT, the serialization of variable-sized data (such as branches containing C-style arrays) produces two internal arrays: one contains the branch data for each event while the other contains the byte offset of each event in the branch data.  For example, if each entry in a branch representing a \texttt{char} arrays contains precisely one entry, then the offset array will contain the integer sequence \texttt{1, 2, 3, 4, ...}.  \textit{LZ4} achieves its performance by looking for byte-aligned patterns (as opposed to \textit{ZLIB}, which works on individual bits) and lacks the Huffman encoding pass; this results in the offset array sequence being effectively incompressible using LZ4.

To improve the performance of LZ4 in this case, we investigated the combination of LZ4 with various ``pre-conditioners".  Pre-conditioners transform the sequence of input bytes according to a simple, deterministic algorithm prior to applying the compression algorithm.  The two algorithms investigated, inspired by the Blosc library \cite{blosc}, are \textit{Shuffle} and \textit{BitShuffle}.  Both preconditioners rearrange the input array's bytes by reading through the array using fixed strides.  For example, if there are 8 bytes in the offset array and the \textit{Shuffle} algorithm uses a stride of 4, the preconditioner's output will shuffle bytes at positions \texttt{1, 2, 3, 4, 5, 6, 7, 8} to \texttt{1, 5, 2, 6, 3, 7, 4, 8}.  If the byte values in the array are \texttt{0, 0, 0, 1, 0, 0, 0, 2} (a big-endian serialization of the 32-bit integers $1$ and $2$), then the preconditioner output would be \texttt{0, 0, 0, 0, 0, 0, 1, 2}.  Given the construction of the offset array, most serialized integers only differ in value by a single byte. The resulting output of the preconditioner often contains long runs of repeated bytes, improving the compression ratio for \textit{LZ4}.

\begin{figure}
\centering
\begin{tikzpicture}[scale=0.75]
\begin{axis}[
    ybar,
    enlargelimits=1.99,
    legend style={at={(0.5,-0.15)},
      anchor=north,legend columns=-1},
    ylabel={Size of files, MB},
    symbolic x coords={nanoaod},
    xtick=data,
    nodes near coords,
    nodes near coords align={vertical},
    ]
\addplot coordinates {(nanoaod,112)};
\addplot coordinates {(nanoaod,79.6)};
\addplot coordinates {(nanoaod,93.2)};
\legend{lz4,lz4-bitshuffle,zlib}
\end{axis}
\end{tikzpicture}
\caption{Comparison of compression ratio in a NanoAOD file from the \textit{LZ4} compressor, \textit{LZ4} with the BitShuffle preconditioner, and the \textit{ZLIB} algorithm.} \label{fig:bitshuffle}
\end{figure}
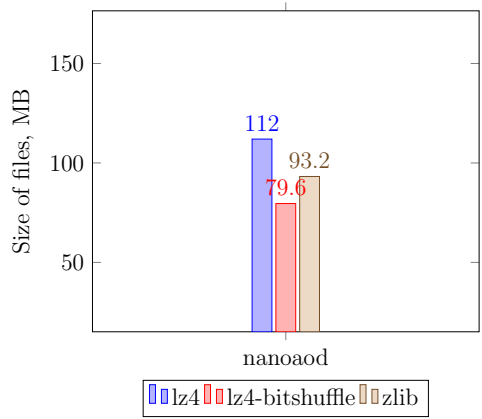

Results from using a preconditioner are shown in Figure \ref{fig:bitshuffle}. The figure shows that, for the sample file (based on the CMS NanoAOD format), the use of the BitShuffle preconditioner results in a variant of the \textit{LZ4} compressor outperforming \textit{ZLIB} in compression ratio.  This variant maintains the significant advantage in decompression speeds from use of \textit{LZ4}.

\subsection{ZSTD}

A newer compression algorithm is ``Zstandard" (or \textit{ZSTD}) \cite{zstd} \cite{zstd-facebook} which aims to replace their existing usage of \textit{ZLIB} with a compression algorithm containing a modern design, aiming to improve compression ratio, compression speed, and decompression speed with respect to \textit{ZLIB}.  The \textit{ZSTD} algorithm is becoming widely adopted in both open-source frameworks (such as Hadoop) and by technology such as Facebook. In addition to the streaming compression engine, interesting features offered by \textit{ZSTD} are bigger window size, and support for generating compression dictionaries.

The window size for \textit{ZSTD} is 256 KB, eight times larger than the \textit{ZLIB} window.  As \textit{ZSTD} is able to search a larger history for pattern matches, it is able to achieve a higher compression ratio.
Further, \textit{ZSTD} provides an additional passs - Finite State Encoding - that outperforms \textit{ZLIB}'s Huffman coding pass in terms of compression ratio and speed.
In addition to these algorithmic changes, the modern design of the library itself allows it to work well with modern processors and compilers. Taken together, we believe these improvements could be advantageous for the data and storage management of \textit{LHC} experiments for Run 3.

Another headline feature of the \textit{ZSTD} is dictionary generation. A dictionary - an array of commonly-used byte sequences in the input - can be provided to the compression algorithm's, allowing the compressor to start from specified state as opposed to building a new one from scratch.  This prebuilt ``lookup table" increases the compression ratio, particularly when compressing a small amount of data (such as a few hundred bytes). The dictionary is generated by analyzing sample data, under the assumption that later data will have similar patterns.  The \textit{ZSTD} library assists in the dictionary generation from a given corpus; however, it does not provide input on the sizing of dictionary or how to effectively store it within the ROOT file.


Performance results for a preliminary integration of \textit{ZSTD} in ROOT were included in Figures \ref{fig:compression} and Figure \ref{fig:decompression}; we aim to include \textit{ZSTD} support in ROOT 6.20.00.

\section{Results}
As part of our initiative to improve ROOT-based analysis, we have continued the work started in \cite{brianzhe} on comparing compression algorithms performance. We included a detailed overview of the advantages and disadvantages of  available compression algorithms integrated in ROOT, such as LZ4, ZSTD, ZLIB and ZLIB-CF. Results of these investigations can be utilized by HEP experiments in order to tune their configurations of ROOT I/O.

Future work items in this program of work include integration of \textit{LZ4} preconditioners in a ROOT release, potentially allowing that algorithm to be used by default for analysis use cases (which we believe are less sensitive to compression ratio but highly sensitive on decompression speed).  We are encouraged by the \textit{ZSTD} performance results and believe it might be a replacement of \textit{ZLIB} for general purpose work.  As the number of supported algorithms in ROOT increases, we believe improvements are needed to the I/O APIs to ease the switch between compression algorithms and settings for different use cases.  The relative ROOT I/O performance results for varying algorithms should also be automatically tracked using the performance benchmarking suite for ROOT \cite{rootbench}.  Finally, we believe the dictionary generation found in the \textit{ZSTD} could provide significant gains in compression ratios; while \textit{ZSTD} can be used to generate the dictionary, the generated dictionaries are useable for \textit{ZLIB} and \textit{LZ4} as well.  Work, however, is needed, to better understand the optimal dictionary sizes and placement within the ROOT file.

\section{Acknowledgments}

This work has been supported by U.S. National Science Foundation grant ACI-1450323 (DIANA-HEP).  The hardware for the ARM tests were graciously provided by CERN Techlab.

\end{document}